\documentclass{epl}

\title{Efficient creation of molecules from a cesium Bose-Einstein condensate}
\shorttitle{Molecules}
\author{M. Mark\inst{1} \and T. Kraemer\inst{1} \and J. Herbig\inst{1} \and C. Chin\inst{1} \and H.-C. N\"{a}gerl\inst{1} \and R. Grimm\inst{1,2}}
\institute{
  \inst{1} Institut f\"{u}r Experimentalphysik, Universit\"{a}t Innsbruck, Technikerstra\ss e 25, 6020 Innsbruck, Austria\\
  \inst{2} Institut f\"{u}r Quantenoptik und Quanteninformation, \"{O}sterreichische Akademie der Wissenschaften, 6020 Innsbruck, Austria }


\begin{document}

\maketitle

\begin{abstract}
We report a new scheme to create weakly bound Cs$_2$ molecules
from an atomic Bose-Einstein condensate. The method is based on
switching the magnetic field to a narrow Feshbach resonance and
yields a high atom-molecule conversion efficiency of more than
30\%, a factor of three higher than obtained with conventional
magnetic-field ramps. The Cs$_2$ molecules are created in a single
$g$-wave rotational quantum state. The observed dependence of the
conversion efficiency on the magnetic field and atom density shows
scattering processes beyond two-body coupling to occur in the
vicinity of the Feshbach resonance.
\end{abstract}


The newly emerging field of quantum-degenerate molecules provides
intriguing possibilities for controlled studies of multicomponent
matter-wave systems. Chemical reactions are expected to show
effects of coherence, matter-wave interference, quantum tunneling,
and bosonic stimulation. Recently, coherent atom-molecule
couplings \cite{amcoherence} have been observed in a Bose-Einstein
condensate (BEC) and molecular quantum gases \cite{csmol, namol}
and molecular BECs \cite{mbec} have been realized. The key
ingredient in these experiments has been the presence of
magnetically induced Feshbach resonances \cite{feshbach}. These
resonances provide the variable coupling between atoms and
molecules as a function of an external magnetic field and allow
the conversion of atoms to molecules and vice versa.

In the previous experiments on the creation of ultracold Cs$_2$,
Na$_2$ and Rb$_2$ molecules from the corresponding atomic BECs
\cite{csmol, namol, rbmol} the molecules are formed by ramping the
magnetic field through a Feshbach resonance; see illustration in
Fig.~1. It is expected that during the ramping process the ground
state atom population in the trap is adiabatically and efficiently
converted into molecules in a weakly bound state \cite{adiabatic}.
However, the reported efficiencies using this method are
relatively low: Typically $5\%\sim10\%$ are observed, whereas
$50\%$ to $70\%$ of the atoms are lost during the ramping process.
The missing fraction, the lost atoms which are not converted into
weakly bound molecules, is generally believed to result from the
creation of molecules in states which cannot be detected by the
conventional imaging method, or to consist of ``hot" atoms which
quickly leave the trap \cite{jin, csmol}.

In this paper, we report a high atom-molecule conversion
efficiency in excess of $30\%$ from an atomic BEC based on a novel
switching scheme. This scheme is illustrated in Fig.~1. The
magnetic field is quickly switched from an off-resonant value
$B_{\rm{start}}$ to a field $B_{\rm{test}}$, near the resonance
position $B_{\rm{res}}$. After a variable hold time
$t_{\rm{hold}}$, the magnetic field is quickly lowered well below
the resonance $B_{\rm{end}}$, where atoms and molecules decouple
and can be independently measured. Our new scheme works for
initial magnetic fields $B_{\rm{start}}$ both well above or well
below the resonance. In the latter case, the creation of molecules
cannot be explained in terms of the two-body adiabatic conversion
picture \cite{adiabatic}. An investigation on the atom loss and
molecule creation efficiencies suggests that different scattering
processes are involved near the narrow Feshbach resonance.

\begin{figure}
\onefigure[width=3.2in]{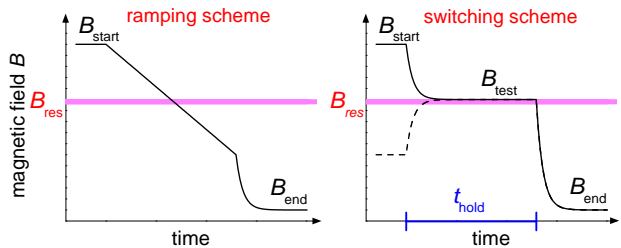} \caption{Schemes for molecule
creation near a Feshbach resonance, located at $B_{\rm{res}}$. In
the ramping scheme, we linearly ramp the magnetic field from
$B_{\rm{start}}$, well above the resonance, to $B_{\rm{end}}$,
well below the resonance. In the switching scheme, we first switch
from $B_{\rm{start}}$ to $B_{\rm{test}}$. After a hold time
$t_{\rm{hold}}$, we switch the field to $B_{\rm{end}}$. The
switching scheme also works for $B_{\rm{start}}$ below the
resonance (dashed line). The finite response time of the magnetic
field in our experiment is due to eddy currents in the stainless
steel vacuum chamber.} \label{fig1}
\end{figure}

The cesium molecules we create are of special interest since they
have a large orbital angular momentum $(l=4)$. Coupling from
ultracold atoms in an $s$-wave scattering state to the $g$-wave
molecular states is observed only for cesium atoms due to the
large indirect spin-spin coupling \cite{spinspin}. Many narrow
Feshbach resonances of this kind were observed at low magnetic
fields for cesium atoms polarized in the lowest internal state
$|F=3,m_F=3\rangle$ \cite{csfeshspec}, where $F$ is the total
angular momentum and $m_F$ is the magnetic quantum number. Based
on these narrow resonances, the formation of thermal molecules was
investigated \cite{chi03} and a pure molecular quantum gas was
created from an atomic BEC \cite{csmol}.

Our experiments start with a pure BEC of cesium with up to
$2.2\times 10^5$ atoms in the ground state $|F=3, m_F=3\rangle$
\cite{csbec, hcnpaper}. The magnetic field is set to 21G,
corresponding to an atomic scattering length of $210a_0$, where
$a_0$ is the Bohr radius. The magnetic field gradient is set to
31.3G/cm for levitation of the atoms \cite{csbec}. The condensate
is confined in a crossed dipole trap formed by two horizontally
intersecting laser beams, which are derived from a broad-band Yb
fiber laser at 1064nm. One tightly focused beam with a waist of
$35\mu$m and a power of 0.5mW essentially provides the radial
confinement; the other beam with a waist of $300\mu$m and power of
350mW essentially provides the axial confinement. The radial and
axial trap frequencies are $\omega_r/2\pi=17.5$Hz and
$\omega_z/2\pi=4.7$Hz, respectively. The chemical potential is
$k_B\times 11$nK, where $k_B$ is Boltzmann's constant.

\begin{figure}
\onefigure[width=3.2in]{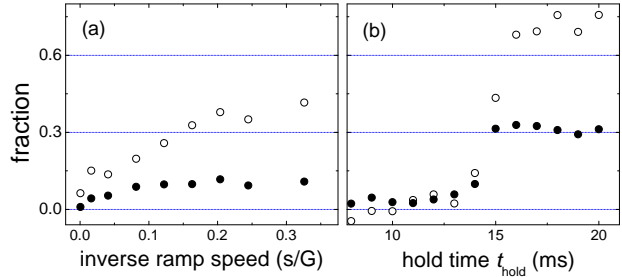} \caption{Comparison of
the two schemes of molecule creation. Molecule fraction (solid
circles) and atom loss fraction (open circles) are shown for (a)
the ramping scheme, where the fractions are measured for different
ramp speeds, and for (b) the switching scheme for different hold
times $t_{\rm{hold}}$. In (b), $B_{\rm{test}}$ is set right on
resonance.} \label{fig2}
\end{figure}

Before we start the molecule creation, we first compress the
condensate by adiabatically increasing the power of the tightly
focused laser in 0.7s to 40mW. The higher laser power provides a
stronger trapping force and allows us to turn off the levitation
field. The absence of the magnetic field gradient is crucial to
ensure that all atoms experience the same magnetic field and can
simultaneously participate in the molecule formation process. In
the compressed trap, the trap frequencies are
$\omega_r/2\pi=170$Hz and $\omega_z/2\pi=6.5$Hz, the chemical
potential is $k_B\times 86$nK and the peak density is $1.7\times
10^{14}$cm$^{-3}$. We then slowly change the magnetic field in
200ms to a starting value of $B_{\rm{start}}$, typically $0.5$G
above the Feshbach resonance $B_{\rm{res}}$. Note that this $0.5$G
offset is much larger than the resonance width of a few mG. The
condensate at $B_{\rm{start}}$ is not influenced by the resonance.
We then switch off the dipole trap and release the atoms into free
space and, at the same time, tune the magnetic field toward the
Feshbach resonance to create molecules. At the end of the molecule
formation phase, we quickly lower the magnetic field down to
$B_{\rm{end}}\approx18$G to decouple the molecules and atoms.

The resulting molecule and atom numbers can be determined
independently by absorption imaging \cite{csmol}. The atoms are
directly imaged at 18G. We verify that the molecules are
insensitive to the imaging beam at this magnetic field. To detect
the molecules, we first blast away the atoms at 18G with a
resonant beam \cite{namol}, and then ramp the magnetic field back
above the resonance to 21G. The weakly bound molecular state is
then above the continuum and the molecules quickly dissociate into
free atoms \cite{disso}. By imaging the cloud of the resulting
atoms, we can determine the molecule number.

\begin{figure}
\onefigure[width=3.2in]{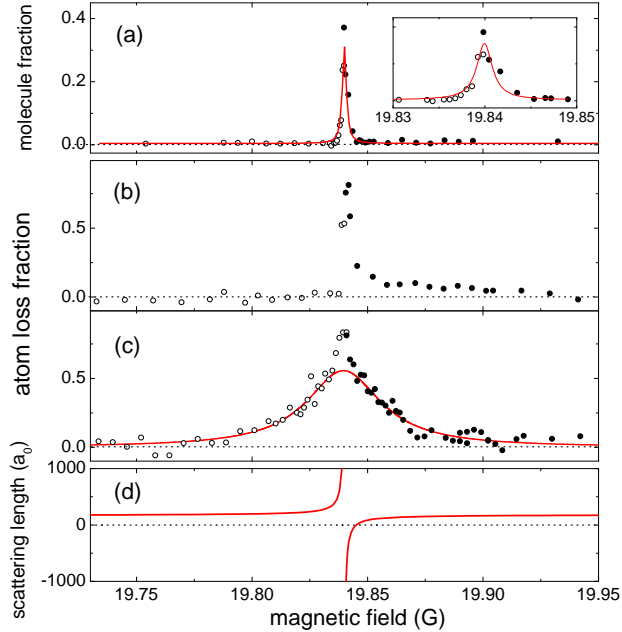} \caption{Molecule creation and
atom loss near the Feshbach resonance. Based on the switching
scheme, molecule increase in $t_{hold}=$ 18ms (a) and atom loss in
18ms (b) are measured for various test fields $B_{\rm{test}}$. The
inset shows the expanded view of the molecule signal. For
comparison, the atom trap loss in the compressed trap is shown in
(c). The scattering length is shown in (d) for reference. Solid
circles (open circles) show the measurements above (below) the
resonance. In (a), a Lorentzian fit (solid line) yields a width of
$2.1$mG and the resonance position of $B_{\rm{res}}=19.840$G,
subject to a calibration uncertainty of 4mG. Fitting both wings in
(c) gives a Lorentzian width of $40(2)$mG.} \label{fig3}
\end{figure}

We employ both the ramping scheme and the switching scheme for
molecule production (see Fig.~1) and compare their performances.
In the ramping scheme, we tune the magnetic field across the
resonance with a constant ramping speed. Based on the resulting
molecule number and the loss in atomic number, we calculate the
conversion fractions, shown in Fig.~2(a). A maximum molecule
fraction of $10\%$ is observed when the ramps are slower than
10G/s. The atom loss for these ramps is about $40\%$, which
indicates a missing fraction of about $30\%$. This result is
similar to all previous experiments using the same method
\cite{csmol, rbmol, namol}. For the switching scheme, we quickly
tune the magnetic field onto the Feshbach resonance
$B_{\rm{test}}\approx B_{\rm{res}}$, wait for various hold times
$t_{\rm{hold}}$, and quickly lower the magnetic field to
$B_{\rm{end}}=18$G. Due to the finite response time of the
magnetic field, the field approaches the Feshbach resonance after
about 12ms \cite{responsetime}. At this time, the peak density of
the expanding condensate reduces to $1.1\times 10^{12}$ cm$^{-3}$
\cite{moleexp}. For hold times $t_{\rm{hold}}>15$ms, molecule
fractions of $30\sim35\%$ and atom loss fraction of $\sim70\%$ are
reached. The conversion efficiency is by more than a factor of
three higher than obtained from the ramping scheme. Note that in
order to precisely set the magnetic field right on the narrow
Feshbach resonance, we synchronize the experiment with the $50$Hz
line voltage to reduce the effects of the ambient magnetic field
ripple, for which we measure an amplitude of 4mG. This suppresses
uncontrolled magnetic field variations to about 1mG.

To understand the different performance of the two schemes, we
study the atom loss and molecule increase at different magnetic
fields $B_{\rm{test}}$ based on the switching scheme, see
Fig.~3(a) and (b). For comparison, we also show the atom loss in
the compressed trap in Fig.~3(c), where the initial peak density
is $1.9\times 10^{14}$ cm$^{-3}$ \cite{density}. The calculated
scattering length is shown in Fig.~3(d) \cite{eite}. For all
measurements with $B_{\rm{test}}$ above the resonance, we prepare
the condensate as previously described at
$B_{\rm{start}}=B_{\rm{res}}+0.5$G. For $B_{\rm{test}}$ below the
resonance, we prepare the condensate at a magnetic field below the
resonance by quickly switching the magnetic field from the initial
value to $B_{\rm{res}}-0.5$G. No appreciable atom loss, molecule
formation or condensate excitation is observed in this process. We
then follow the same experimental procedure, but approach the
resonance from below. These two different preparation procedures
for magnetic fields above and below the resonance are necessary to
avoid a slow field-sweep across the resonance, which can lead to
systematic atom loss or molecule increase.

In the molecule creation spectrum (Fig.~3 (a)) we observe a very
narrow linewidth of 2.1mG, which is consistent with the predicted
resonance width. Notably, our molecule creation scheme also works
for $B_{\rm{start}}$ below the resonance, which suggests that
coupling beyond the adiabatic conversion model plays an important
role in the creation process. In the adiabatic passage picture,
molecules cannot be created when the creation field is below the
resonance. The atom loss, shown in Fig.~3(b), is asymmetric and
seems to include two components, a narrow peak on resonance and a
much broader and weaker loss feature for magnetic fields above the
resonance. The narrow peak has a similar width as in the molecule
production spectrum in Fig.~3(a), and is clearly related to the
observed molecule formation. The broad and weak feature on the
high magnetic field side has a width of 80(20)mG as determined
from a one-sided Lorentzian fit. To obtain further information
about the atom loss process, we measure the atom loss in the
compressed trap, where the atom density is higher by a a factor of
$\sim 170$ than in Fig.~3(a) and (b). The result shown in
Fig.~3(c) displays a wide and symmetric loss feature. By fitting
the two wings to a Lorentzian profile, we find a width of 40(2)mG.

The different lineshapes suggest that different scattering
processes are involved near the Feshbach resonance. The molecule
formation width is close to the predicted width of the Feshbach
resonance and can be interpreted in terms of the two-body Feshbach
coupling. The asymmetric loss feature in Fig.~3(b) and the trap
loss may be due to three-body recombination or many-body effects.
These broad atom loss features are puzzling, since they are a
factor of $20$ or more wider than the Feshbach resonance width of
2mG. The physical origin of the associated loss mechanisms
requires further investigation.

The large width of the atom loss feature, however, does provide a
qualitative explanation why the switching scheme is more efficient
than the ramping scheme. In a linear ramp, atoms sample all
magnetic fields near the resonance which, for a large fraction of
time, leads to atom loss without molecule increase. With the
switching scheme, the atoms spend more time in the magnetic field
range where the molecules can be created.

\begin{figure}
\onefigure[width=3.2in]{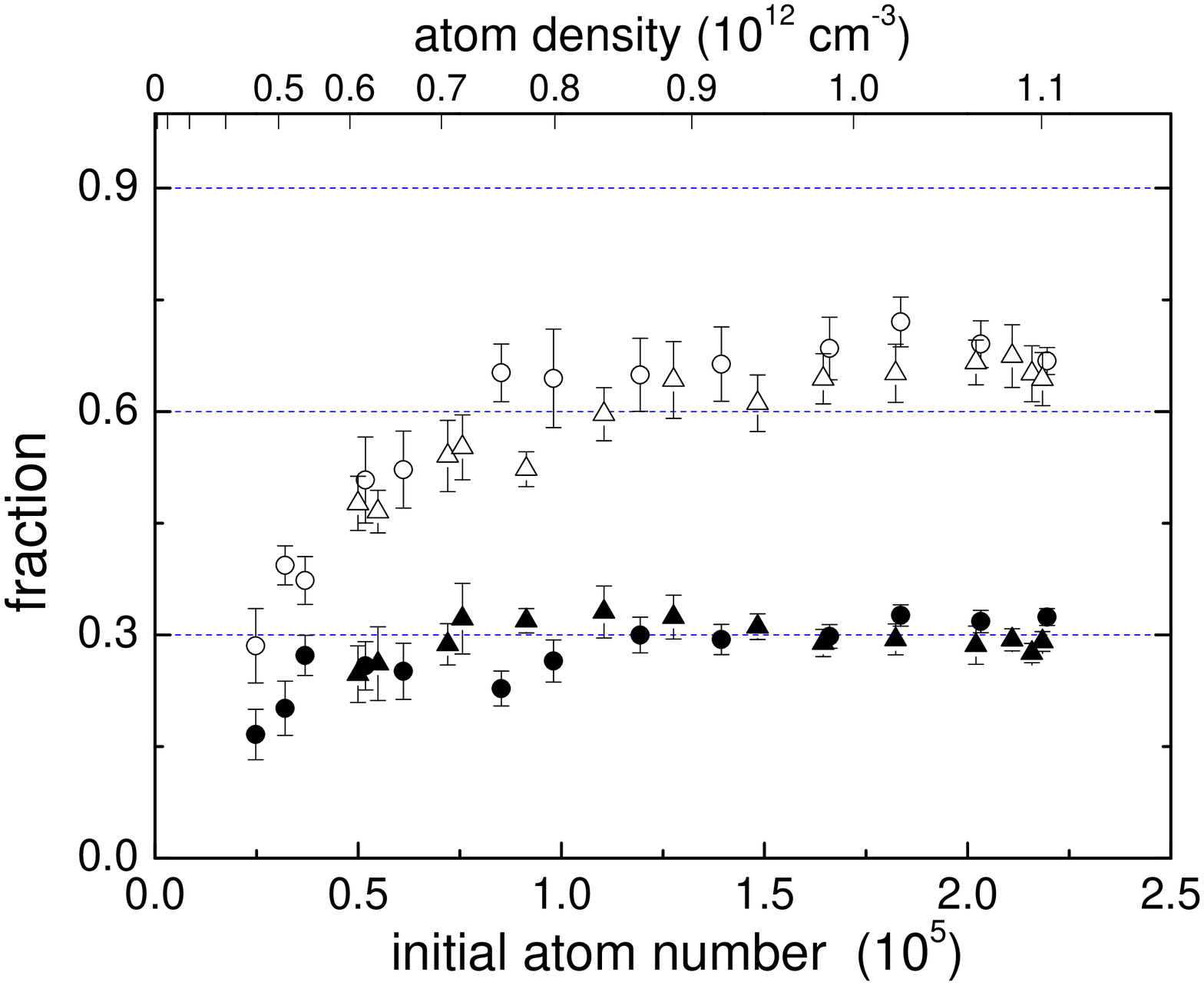} \caption{Density dependence of
the atom loss fraction (open symbols) and molecule fraction (solid
symbols). The atom number in the BEC is varied by either adjusting
optical cooling efficiencies (solid and open circles) or the
loading efficiencies into the crossed dipole trap (solid and open
triangles). The corresponding atomic density of the sample right
before the molecule formation (after 12ms expansion in free space)
is given on the top axis.} \label{fig4}
\end{figure}

To further investigate and differentiate the physical mechanisms
that are responsible for the missing fraction and for the molecule
increase, we measure the dependence of the atom loss fraction and
molecule fraction on the atom number in the condensate, as shown
in Fig.~4. Atom numbers ranging from $2.5\times10^4$ to $2.2\times
10^5$ correspond to peak densities of $7.9\times 10^{13}$cm$^{-3}$
to $1.9\times 10^{14}$cm$^{-3}$ in the compressed trap and to
$4.6\times 10^{11}$cm$^{-3}$ to $1.1\times 10^{12}$cm$^{-3}$ in
free space at the moment the molecules are created. For the
calculation of the densities in free space, we take into account
the $12$ms expansion of the condensate in the Thomas-Fermi regime
after the dipole trap is turned off.

Several interesting features show up in the density dependence.
The molecule fraction grows and saturates to $\sim 30 \%$ at
densities higher than $5\times 10^{11}$cm$^{-3}$.  The saturation
of the molecule fraction resembles observations in a thermal gas
\cite{chi03, selim1}, where a thermal equilibrium is reached with
a constant molecule fraction in the sample \cite{atmoleq}. The
missing fraction is very small at low densities and continues to
grow up to a density of $8\times 10^{12}$cm$^{-3}$. The stronger
density dependence of the missing fraction suggests that
scattering processes involved in the atom loss are of higher order
than for the molecule increase. Similar enhancement of the
collision loss near the Feshbach resonance was also observed in a
$^{85}$Rb condensate \cite{rb85} and in a thermal Cs gas
\cite{tino}. A further analysis on the scattering dynamics and the
possible thermal equilibrium condition is necessary.

In conclusion, we show that an atom-molecule conversion fraction
of more than $30\%$ can be reached based on a magnetic field
switching scheme. The performance of this scheme is superior to
the conventional linear magnetic field ramping scheme since the
molecules are created only within the narrow Feshbach resonance
width of 2mG, while the atom are lost over a much large range of
$\sim40$ mG. The density dependence of both the missing fraction
and the molecule fraction suggests that in our scheme the
molecules are created via Feshbach coupling, while the missing
fraction comes from higher order scattering processes. Based on
the new creation scheme, we are now able to obtain samples with up
to 40,000 ultracold molecules. This provides a good starting point
to investigate the trapping, the interactions, and the matter-wave
nature of ultracold molecules.

\acknowledgments We acknowledge support by the Austrian Science
Fund (FWF) within SFB 15 (project part 16) and by the European
Union in the frame of the Cold Molecules TMR Network under
Contract No.\ HPRN-CT-2002-00290. M.M. is supported by DOC
[Doktorandenprogramm der \"{O}sterreichischen Akademie der
Wissenschaften]. C.C.\ is a Lise-Meitner research fellow of the
FWF.


\begin{thebibliography}{0}
\bibitem{responsetime}
The response time of $t_{1/e}$=1.54ms for the magnetic field is
limited by the eddy currents from the vacuum chamber. Starting
from 500mG above the resonance, we estimate the magnetic field
will take $\ln(500$mG$/2$mG$) t_{1/e}\approx 9$ms to settle within
the resonance width. This result is close to the measured delay
time of 12ms.
\bibitem{moleexp}
We observe that the expansion rate of the molecules is very close
to that of the expanding condensate.
\bibitem{density} This value is slightly higher than the value given before
because of the reduced scattering length.
\bibitem{eite}
The calculated resonance position is $20.1$G. To reflect the
correct resonance position, we shift the numerical data by
centering the resonance at our observed value of 19.84G.
\end{thebibliography}
\end{document}